\documentclass[aps,prb,amsmath,amssymb,reprint,author-year,author-numerical,floatfix,superscriptaddress]{revtex4-2}

\usepackage{graphicx}
\graphicspath{{./images/}}
\usepackage{dcolumn}
\usepackage{bm}
\usepackage{mathtools}
\usepackage{amssymb}
\expandafter\let\csname equation*\endcsname\relax
\expandafter\let\csname endequation*\endcsname\relax
\usepackage{amsmath}
\usepackage{amsbsy}
\usepackage{booktabs}
\usepackage{verbatim}
\usepackage{array}
\usepackage{braket}
\newcolumntype{P}[1]{>{\centering\arraybackslash}p{#1}}
\usepackage[colorlinks = true,
            linkcolor = blue,
            urlcolor  = blue,
            citecolor = blue,
            anchorcolor = blue]{hyperref}
\AtBeginDocument{}%

\begin{document}

\preprint{AIP/123-QED}

\title[FISCHETTI]{On the electronic and vibrational dimensionality of nanometer-scale silicon structures}

\author{Massimo V. Fischetti}
\email[email: ]{max.fischetti@utdallas.edu.}
\affiliation{Department of Materials Science and Engineering, The University of Texas at Dallas\\ 
             800 W. Campbell Rd., Richardson, TX 75080}   
\author{Dallin O. Nielsen}
\affiliation{Department of Materials Science and Engineering, The University of Texas at Dallas\\
             800 W. Campbell Rd., Richardson, TX 75080}
\author{Edward Chen}
\affiliation{Corporate Research, Taiwan Semiconductor Manufacturing Company Ltd.\\
             8 168, Park Ave. II, Hsinchu Science Park, Hsinchu 300-75, Taiwan}                                 

\date{\today}

\begin{abstract}
We discuss the problem of assessing the electronic and vibrational dimensionality of a semiconductor nanostructure: How thin and/or wide must a nanostructure be in
order to induce electron and phonon confinement? Clarifying the physical justification for common criteria found in the literature, we view the electron coherence
length (defined as the electron and phonon inelastic mean free path) as their `field of view' and argue (or, better yet, `speculate') that this sets the important
length scale. Considering the example of Si nanosheets at room temperature, and drawing from results found in the literature, we estimate that the critical length 
below which electrons are subject to quantum confinement is of the order of (or smaller than) 8~nm, when their coherence length is determined by energy losses to
phonons and remote phonons in gated structures. On the contrary, no single length-scale can be given for phonons: Taking their coherence length as determined by
scattering with electrons and anharmonic three-phonon processes, short wavelength acoustic and optical phonons may be confined only by structures as small as 10~nm.
Long-wavelength acoustic phonons, instead, may exhibit a coherence length of the order of 1~$\mu$m, so that they may be confined over much larger distances.     
\end{abstract}

\maketitle

\section{\label{sec:Intro}{Introduction}} 
In our recent theoretical study of electron transport in double-gate Si nanosheets 1.6~nm-thick and 12~nm wide~\cite{Mansoori_2026}, we based the entire study on   
the implicit assumption that electrons and phonons are quantized by the geometric confinement along the $z$ axis (along the direction perpendicular to the
interfaces), but not along $y$ axis (the direction on the $(x,y)$ plane along the width of the nanosheet, perpendicular to the transport direction $x$).\\ 

The correctness of these assumptions is far from obvious. Common well-known, 'intuitive' and 'obvious' criteria used to assess whether particles (or quasi-particles)
should be considered `quantized' by the geometric confinement rely on determining whether the length-scale of the confinement is shorter than the particle
mean free path~\cite{Baalousha_2014} (presumably inelastic), of the effective Bohr radius or de Broglie wavelength~\cite{Neikov_2019,Nozik_2021}. Alternatively, 
one may consider the energy separation among the quantized states~\cite{Fischetti_1988,Reaz_2021}: If this separation is smaller than the thermal energy, 
$k_{\rm B}T$, or some other appropriate energy scale, then the discrete nature of the states is thought to be `washed out' by the broad energy distribution of the thermal particles and they are treated 
s bulk, 3D particles.\\

As 'obvious' as these criteria may be, their physical foundation is almost always left to intuition rather than justified quantitatively and the literature shows different authors reaching vastly different conclusions. On the one hand, electrons are often assumed to be 
aubject to quantum confinement even over large distances. For example, this assumption has been made by 
Liu {\it et al.}~\cite{Liu_2024} in Si nanosheets/ribbons 3~nm thick and up to 12~nm-wide; by Donetti {\it et al.}~\cite{Donetti_2025} in Si nanowires with thickness
of 4~nm and a width as large as 30~nm; by Kim {\it et al.}~\cite{KYKim_2025} in nanosheets with a 4$\times$8~nm$^2$ cross section; by Ramayya and coworkers~\cite{Ramayya_2006,Ramayya_2008} for 8~nm-thin Si nanosheets with widths ranging from 8~nm all the way up to 30~nm, and for nanowires with cross 
sections as large as 8$\times$8~nm$^2$; by Islam {\it et al.}~\cite{Islam_2018} even for silicon-on-insulator (SOI) bodies as thick as 8~nm; and in
Refs.~\cite{Luisier_2013},\cite{Dixit_2025}, and \cite{Lee_2026} in Si nanosheets with widths in the 8-to-20~nm range. In these studies the electrons are 
treated as a 1DEG, even considering the largest structures, up to 30~nm. Similarly, in Ref.~\cite{Hattori_2024} electrons have been considered as a confined 
2DEG in 40-nm-wide Si thin films even for a thickness as large as 7~nm. On the other hand, in Refs.~\cite{Thobel_2010} and \cite{Popescu_2015} electrons are
treated as a bulk, 3D electron gas in InAs nanowires with cross section of a few tens of nm, despite the fact that electrons in InAs exhibit a very long mean free 
path.\\

Obviously, assessing the dimensionality of the electron gas
in a channel of a transistor is necessary to understand not only its electrostatic properties (such as the threshold voltage) and the carrier kinematics and dynamics
({\it i.e.}, the group velocity, dispersion, and scattering rates), but also parasitic effects, such as the injection from a bulk source extension into
a 2D or 1D channel, the problem related to quantum point contacts~\cite{Szafer_1989}, recently studied by Kim {\it et al.}~\cite{KYKim_2025} 
in the context of Si nanosheets. This is a pressing issue, 
now that device scaling has forced us to consider smaller and smaller nanostructures as active channels, from ultra-thin-body (UTB) SOIs~\cite{Bokor_1999} thinner
than 20~nm, to finFETs with fins as narrow as 4~nm~\cite{Chang_2011,He_2017}, from nanowires~\cite{Yoon_2020} to Si 
nanosheet/ribbons~\cite{Yun_2006,Lee_2018,Mukesh_2020,Agrawal_2024,Yeap_2024} as thin as 1.6~nm and as narrow as 12~nm.\\

Quite surprisingly, the problem of determining the dimensionality of a system ({\it i.e.}, should we consider electrons as a 3D, a 2D, or even a 1D gas?) seems to have
been overlooked in the literature. A recent article~\cite{Foller_2024} carries a tantalizing title, but it does not discuss when a thin film of a semiconductor 
acquires two-dimensional electronic or vibrational properties, such as the electronic band structure and transport or the phonon spectrum. Therefore, focusing on the
aforementioned Si nanosheets as an example of utmost technological interest, here we intend to justify the 'obvious' physical criteria that we must use to address 
this problem, at least in a qualitative way, and provide some quantitative estimates.\\ 

We shall proceed by looking at the problem in general in Sec.~\ref{sec:Generalities} considering electrons, just to fix the ideas. Our considerations may be obvious, but we think it is worth to state them explicitly, given the confusing status of the literature. We shall then attempt to analyze the problem quantitatively 
considering electrons, in Sec.~\ref{sec:Electrons}, and phonons, in Sec.~\ref{sec:Phonons}, in the specific case of the 1.6~nm-thick, 12~nm-wide Si nanosheet in a double-gate structure with SiO$_2$/HfO$_2$ gate stacks. 

\section{\label{sec:Generalities}{A general qualitative overview of the problem}} 
The common criterion of defining the particle mean free path as the critical length scale below which quantization occurs, should be better specified by defining, more
precisely, such a length scale as given by the magnitude of the electron and phonon phase-coherence (or `dephasing') lengths, $\lambda_{\phi, {\rm el}}$ and 
$\lambda_{\phi, {\rm ph}}$, respectively. These represent the distances over which these particles or quasi-particles travel coherently without losing memory 
of their phases. Such a loss of coherence can be seen in Ref.~\cite{Baek_2016} that discusses the time evolution of the Green's function of a particle in a
`box', showing how the particle may become localized in the subregion of the box as a consequence of a delta-like time-dependent perturbation. 
This coherence length is set by the inelastic scattering processes (that is, by time-dependent perturbations caused by scatterers that possess internal 
degrees of freedom),
such as electron-phonon scattering or anharmonic 3-phonon processes. Elastic scattering, such as electron/ionized-impurity scattering or scattering at a rough
interface, become dissipative processes only after some configuration average; however, in principle, they do not break the phase.\\

Therefore, as an `educated guess', we may consider that particles (electrons) or quasi-particles (phonons) are `quantized' along one direction when their 
phase-coherence length is at as long as the confining length. On the contrary, if the phase-coherence length is shorter than the confining length, the 
(quasi)particles should not be treated as quantized. The `gray' intermediate region ({\it i.e.}, should the coherence length be twice as long as the confining length
to allow for a back-n-forth travel and set up standing waves?)
is hard to treat and we shall leave it as such, `gray'. In simple words, one may view the electron or phonon coherence lengths as their `horizons' 
(or 'lines of sight' or `fields of view'): If the confinement could be potentially caused by a band (elastic or dielectric) discontinuity, the electron (phonons) 
must be able to `see' the discontinuity so that it may be reflected and set up the standing wave the constitutes the bound state. If such a discontinuity, instead, is beyond their 'horizon', they should be described as bulk (quasi)particles, albeit in strongly perturbed and collisionally-broadened bulk states.\\

   A more rigorous analysis would be welcome but, as we said, the problem does not seem to have been rigorously studied in the literature. We can consider here the 
   `toy problem' of an electron of mass $m$ in a potential well with infinitely hard boundaries (that is, a `box' of width $L$) with a time- and space-dependent 
   perturbation $V_q e^{iqz} e^{i \Omega_q t}$ that mimics the electron-phonon interaction.
   To first order, the perturbation induces a scattering (or 'decoherence') rate for an electron in the unperturbed state $\ket{\mu}$ as:   
   \begin{equation}
   \frac{1}{\tau_{\phi, \mu}} = \frac{2 \pi}{\hbar} \sum_{\nu, q} |\braket{\nu|V_q|\mu}|^2\ \delta (E_\nu -E_\mu + \hbar \Omega_q) ,   
   \label{eq:box_4a}
   \end{equation} 
   where $E_\nu$ is the unperturbed energy of the eigenstate $\ket{\nu}$. 
   The state $\ket{\mu}$ will have a scattering-induced broadening $\Delta E_\mu = \hbar/\tau_{\phi, \mu}$. We can still treat 
   the system as a 2DEG as long as the broadening (half-width at half maximum, HWHM) remains a small fraction of the unperturbed energy; that is:
   \begin{equation}
   \Delta E_\mu = \hbar/\tau_{\phi, \mu} < E_\mu = \frac{\hbar^2 \pi^2}{2m L^2}\mu^2 \ .   
   \label{eq:box_5}
   \end{equation}
   Defining the phase-coherence length $\lambda_{\phi, \mu}$ as $\upsilon_{\mu} \tau_{\phi,\mu}$ (where $\upsilon_{\mu}=\hbar \pi \mu/(mL)$ is the electron 
   group velocity in the state $\ket{\mu}$), this equation implies:
   \begin{equation}
   \lambda_{\phi,\mu} > \frac{2}{\mu \pi} L \ .
   \label{eq:box_6}
   \end{equation}
   Therefore, the strictest condition we must satisfy in order to treat the electrons as a 2DEG occurs for the ground-state $\mu=1$ and requires 
   $\lambda_{\phi} > 2L/\pi \sim L$. Being interested only in rough estimates, we ignore factors of 2 that can arise from considering the electron phase velocity 
   or the half-width of the broadened states, and view the condition $\lambda_{\phi} > L$ as the minimum requirement that must be satisfied in order to consider 
   the electrons as confined. If, instead, the broadening approaches or exceeds the magnitude of the unperturbed energy, $\Delta E_\mu \gtrapprox E_\mu$, then 
   the eigenstates of the 2DEG `merge' (the common criterion mentioned above with the broadening arbitrarily set to $k_{\rm B}T$) 
   and we recover a `renormalized' 3D continuous spectrum.\\

   To express these considerations a bit more formally for a real system, the spectral density, $A({\bf K}, \omega)$, of an electron (with masses $m_z$ and
   $m_{\parallel}$ along the    quantization and in-plane directions, respectively) in a ground state of a quantum well, 
   which is just $\delta[\hbar\pi^2/(2m_z) + \hbar K^2/(2m_{\parallel})- \omega]$ in the absence of the perturbation, would be broadened as
   \begin{multline}
   A({\bf K}, \omega) = 
      \frac{1}{\pi} {\mbox{Im}} \left [ \frac{1}{\omega-\omega_0({\bf K}) -\Sigma_0({\bf K},\omega)} \right ] = \\
      \frac{1}{\pi} \frac{|{\mbox{Im}}[\Sigma_0({\bf K},\omega)]|}
                          {[\omega-\omega_{0}({\bf K})-{\mbox{Re}}[\Sigma_0({\bf K},\omega)]]^2 + {\mbox{Im}}\Sigma_0({\bf K},\omega)^2} \ ,
   \label{eq:spectral_density}
   \end{multline}  
   where $\hbar \omega_0({\bf K}) = \hbar \pi^2/(2m_z)+\hbar K^2/(2m_{\parallel})$ is the unperturbed energy of the ground-state and $\Sigma_0({\bf K},\omega)$ 
   is the self-energy of electrons in the ground state. To the leading order in perturbation theory this is given by:
   \begin{multline}
   \Sigma_0({\bf K},\omega) = \lim_{\eta \rightarrow 0^+} \sum_{\mu {\bf Q}} \ |g_{0, \mu}({\bf K},{\bf Q})|^2   \\
     \times \left [
          \frac{N_{\bf Q} + f_{\mu,{\bf K}+{\bf Q}}} {\omega-\omega_{\mu,{\bf K}+{\bf Q}}+\Omega_{\bf {Q}} - i \eta} + 
          \frac{N_{\bf Q} + 1 - f_{\mu,{\bf K}+{\bf Q}}} {\omega-\omega_{\mu,{\bf K}+{\bf Q}}-\Omega_{\bf {Q}} + i \eta} \right ]  \ .
   \label{eq:Sigma_1}
   \end{multline}
   Here $N_{\bf Q}$ and $f_{\mu,{\bf K}}$ are the phonon and electron occupation, respectively, 
   $\omega_{\mu,{\bf K}}=\hbar \pi^2 \mu^2/(2m_z)+\hbar K^2/(2m_{\parallel})$ is the frequency of the unrenormalized electronic states, and 
   $g_{0,\mu}({\bf K},{\bf Q})$ is  the electron-phonon matrix elements for a phonon with wavevector ${\bf Q}$. 
   Thus, $\Sigma_0({\bf K},\omega)$ is the (complex) electron self-energy due to the transitions to states $\mu$ caused 
   by absorption/emission of phonons of momentum ${\bf Q}$ (having ignored the Debye-Waller contribution for simplicity).
   $|{\mbox{Im}}[\Sigma_0({\bf K},\omega)]| \approx (1/2) \hbar/\tau({\bf K})$ is the imaginary part of the electron self-energy representing its lifetime 
   $\tau({\bf K})$, and ${\mbox{Re}}[\Sigma_0({\bf K},\omega)]$ represents the renormalized dispersion.\\
   
   We may now view the electron state as a wave packet 
   constituted by the superposition of a broad range of off-shell states with wavevectors ${\bf K}+{\bf Q}$. 
   This is not too different from the packet that we would have in the absence of the quantum-well, since the electron cannot `see' the walls and its coherence 
   length is similar to a bulk, 3D electron. At the lowest order in perturbation theory, assuming bulk phonons, Eq.~(\ref{eq:Sigma_1}) would differ from its bulk
   expression because of a different dispersion (so, $\omega_{\mu,{\bf K}}$ and $f_{\mu,{\bf K}}$) and the different wavefunctions (bulk instead of confined) that 
   appear in the matrix elements $g_{0,\mu}({\bf K},{\bf Q})$. 
   For a strong electron-phonon coupling, the self-energy $\Sigma_0({\bf K},\omega)$ (and, so, the spectral density $A({\bf K},\omega)$) would be controlled 
   mainly by $|g_{0,\mu}({\bf K},{\bf Q})|$ and this expression would approach the same form it would take in the 
   absence of the confining potential well: The broadening would result in a continuum spectrum, the index $\mu$ being replaced by the continuum in the direction of
   the confinement. In other words, the density of states $\rho(E)$, given in terms of the spectral density, 
    \begin{equation}
   \rho(E) = \frac{2}{(2 \pi)^2} \int {\rm d} {\bf K} \ A({\bf K}, E/\hbar) \ ,
   \label{DoS_spectral}
   \end{equation}
   would also be similar to the DoS of a strongly perturbed (renormalized) 3DEG.
   In simple terms, the electron excitation spectrum in the quantum well would be indistinguishable from the excitation spectrum in the absence of the well;
   electron-phonon scattering prevents the electrons from 'seeing' the walls and their dispersion approaches the bulk dispersion. So, the original
   heuristic criterion we mentioned before, that quantum confinement can be ignored when the energetic separation of the confined eigenstates is smaller than the 
   thermal energy, $k_{\rm B}T$, is correct provided $k_{\rm B}T$ is replaced by the collisional broadening ${\rm Im}[\Sigma_0({\bf K},\omega)]$.  \\
   
   Getting back to Eqs.~(\ref{eq:box_5}) and (\ref{eq:box_6}), since $\upsilon_{\mu}$ is the component of the group velocity along 
   the quantization direction, the frequency, $f^{\rm (hit)}_{\mu}$ at which the electron hits the confining potential `walls' is $\upsilon_{\mu}/L$ 
   and Eq.~(\ref{eq:box_6}) can be restated simply as: 
   \begin{equation}
   2  {\rm Im}[\Sigma_0(\mu)]/\hbar = \frac{1}{\tau_{\phi,\mu}} < \frac{\pi \mu}{2} \frac{\upsilon_{\mu}}{L} \sim f^{\rm (hit)}_{\mu} \ , 
   \label{eq:box_7}
   \end{equation} 
   implying that the electron states are quantized when the frequency at which the electron hits the walls is larger than the decoherence rate.\\
     
   Finally, is $\lambda_{\phi}$ the coherence length of a 2DEG (or whatever is the lower-dimensionality of the confined electrons), $\lambda_{\rm \phi 2D}$, 
   or of bulk electrons (or whatever is the higher dimensionality of the `unconfined' electrons), $\lambda_{\rm \phi, bulk}$? If the condition
   $\lambda_{\rm \phi, 2D}, \lambda_{\rm \phi, bulk} \gg L$ is satisfied, then the electrons are certainly confined. Similarly, if 
   $\lambda_{\rm \phi, 2D}, \lambda_{\rm \phi, bulk} \ll L$, then the electrons remain three-dimensional. Although these are the conditions
   considered below, in principle one may envision situations in which $\lambda_{\rm \phi, bulk} \gtrapprox  L$ but $\lambda_{\rm \phi, 2D} \lessapprox L$ (as it may 
   be when the 2D scattering rates are larger than the bulk rates, as a result of quantum confinement) or, instead 
   $\lambda_{\rm \phi, 2D} \gtrapprox  L$ whereas $\lambda_{\rm \phi, bulk} \lessapprox L$ (as it may be when the 2D scattering rates are smaller than the bulk rates, 
   as a result of intervalley scattering present in 3D but not in 2D, as a result of the large energy separation between excited and satellite subbands). 
   It is sensible to think that, in order to have a lower dimensional electron gas, the confined system must be internally consistent; that is, 
   the condition $\lambda_{\rm \phi, 2D} > L$ must always be satisfied. Indeed, considering quantum confinement as a dynamic process, an electron `introduced' 
   in a quantum well will attempt to set up a standing wave. Regardless of whether $\lambda_{\rm \phi, bulk}$ is larger or smaller than $L$, eventually the electron
   stochastically will have the chance to retain memory of its phase for a time long enough to set up such a wave. If $\lambda_{\rm \phi, 2D} >  L$, as soon as this 
   happens, the electron will remain permanently `quantized'. If, instead, $\lambda_{\rm \phi, 2D} < L$, the standing wave will dissipate, again regardless of whether 
   $\lambda_{\rm \phi, bulk}$ is larger or smaller than $L$.\\ 
     
   Admittedly, these are just conjectures or, better, `educated guesses'. 
   Presumably, the `correct' electronic excitation spectrum should be obtained from self-consistent calculations 
   based on the non equilibrium Green's function (NEGF) method and density functional theory (or even GW) that account for full inelastic electron-phonon
   interactions, a rather daunting task. Nevertheless, we believe that our considerations capture the major physical ingredients of a problem that, as obvious 
   as it may be, is often taken as granted, or even completely overlooked, in the literature.   
\section{\label{sec:Electrons}{Electron confinement (and transport) in S\lowercase{i} nanosheets}}
\subsection{Electron confinement} 
Moving to the specific case of Si nanosheets to obtain some quantitative estimates, we start by considering electron confinement. Their coherence length can be estimated from the total
electron-phonon scattering rate, $1/\tau^{\rm (ep)}$ shown in Fig.~7 of Ref.~\cite{Mansoori_2026}. For thermal electrons, 
$\lambda_{\phi, {\rm el}} \sim \upsilon_{\rm g} \tau^{\rm (ep)}_{\rm th}$, where $\upsilon_{\rm g}$ is the electron group velocity along the quantization 
direction and  $\tau^{\rm (ep)}_{\rm th}$ is the electron-phonon scattering time at the thermal energy $\sim k_{\rm B}T$ in two dimensions.\\

From Fig.~6 of Ref.~\cite{Mansoori_2026} we see that $1/\tau^{\rm (ep)}$ ranges from $3 \times 10^{12}$/s to $8 \times 10^{13}$/s, for phonon clamped boundary
conditions (CBCs) at the SiO$_2$/Si interfaces and phonon free-standing boundary conditions (FSBCs), respectively and, from Fig.~7, 
$1/\tau^{\rm (ep)} \approx 3 \times 10^{13}$/s assuming CBCs at the 
SiO$_2$/HfO$_2$ interfaces. Considering these more realistic CBCs and assuming $\upsilon_{\rm g} = \hbar \pi/(m_{\rm L} t_{\rm s})\sim 2.5 \times 10^7$~cm/sec
(the group velocity along the quantization direction for an electron in the ground-state `unprimed' subband), we have 
$\lambda_{\phi, {\rm el}} \sim$~3-to-8~nm assuming $m_{\rm L} \approx 0.91 \ m_0$ and $t_{\rm s} \approx 1.6$~nm. Additional inelastic scattering processes 
(`remote phonon' or `IPP scattering') and, most important, short-range electron-electron Coulomb interactions) may reduce this value. 
Therefore, it appears that for FSBCs, and even 
more so for CBCs, $\lambda_{\phi, {\rm el}} > t_{\rm s}$, so electrons have ample time to `bounce' coherently back-n-forth along the thickness of the nanosheet 
and give raise to the standing waves that amount to quantization along the $z$ axis. This leads us to treat the electrons as a 2DEG.
If, instead of the 2D scattering rates we were to consider those calculated for bulk (3D) electrons with bulk phonons, we would obtain similar or even larger 
dephasing length $\lambda_{\phi}$, since the electron-phonon scattering rates for thermal electrons in bulk Si are of the order of $10^{13}$/s (see Fig. 8 of
Ref.~\cite{Yang_2024}) and that in the electrical quantum limit (so, in films thinner than about 5~nm in which inter-subband and inter-valley scattering are 
not active, only intrasubband scattering with acoustic phonons being permitted -- see below), the rates are even smaller, of the order of $10^{12}$/s (see 
Fig. 17 of Ref.~\cite{Jacoboni_1983}.  This reinforces the conclusion that electrons are indeed a 2DEG in 1.6~nm-thick (100) Si sheets.\\ 

A rough estimate of the critical thickness beyond which we should treat electrons as a bulk 3D particles can be obtained assuming the electrical quantum limit 
(that is: only the ground-state unprimed subband is occupied and only intra-subband scattering is active). In this case, the electron-phonon scattering rates
scale with the inverse of the sheet thickness, as $1/t_{\rm s}$. Writing $\tau_{\phi} = \tau_{\phi,0} t_{\rm s}$, for $1/\tau_{\phi} = 3 \times 10^{13}$/s, 
we have $\tau_{\phi,0} \approx 2 \times 10^{-5}$~s/m. The critical thickness, $t_{\rm s, crit}$, will be given by:
\begin{equation}
\frac{1}{\tau_{\phi,0} t_{\rm s, crit}} \approx f^{\rm (hit)} = \frac{\hbar \pi}{m_{\rm L}t^{2}_{\rm s, crit}} 
\nonumber
\vspace*{-0.20cm}
\end{equation} 
or
\vspace*{-0.20cm}
\begin{equation}
t_{\rm s, crit} = \frac{\hbar \pi}{m_{\rm L}} \tau_{\phi,0} \sim 8 \ \mbox{nm} \ . 
\label{eq:ts-crit}
\end{equation}
This is only a very rough estimate since, when accounting for the confinement of the phonons, $\tau_{\phi}$ increases very slowly (if at all) with decreasing 
$t_{\rm s}$ as the energy of the acoustic phonons at the $\overline{\Gamma}$ point also grows as $1/t_{\rm s}$, resulting in lower scattering rates. 
Moreover, for the unprimed subbands, the electrical quantum limit is valid only for a thickness of about 7-to-8~nm, corresponding to a separation of about 
$k_{\rm B}T \approx$ 25~meV between the ground-state subband and the first excited subband; beyond this, inter-subband scattering and intervalley scattering 
increase the scattering rates. This estimate can be obtained by considering the energy separation between the ground-state and the first excited state of the 
unprimed ($\Delta_2$) doubly-degenerate subbands, $\Delta E_{1,2}$, and the energy separation between the ground states of the unprimed ($\Delta_2$) double degenerate subbands and of the fourfold degenerate primed ($\Delta_4$) subbands, $\Delta E'_{1,1}$. Assuming a simple `particle in the box' model, for Si the
thickness (or width) $t_{\rm s, crit}$ corresponding to $\Delta E_{1,2} \approx \Delta E'_{1,1} \approx k_{\rm B}T$ are
\begin{equation}
t_{\rm s, crit} = \hbar \pi \left [ \frac{3}{2 \Delta E_{1,2}} \frac{1}{m_{\rm L}} \right ]^{1/2} \approx 7 \mbox { nm} 
\nonumber
\end{equation}
and
\begin{equation}
t_{\rm s, crit} = \hbar \pi \left [ \frac{1}{2 \Delta E'_{1,1}} \left ( \frac{1}{m_{\rm T}}-\frac{1}{m_{\rm L}} \right ) \right ]^{1/2} \approx 8 \mbox { nm} \ .
\label{eq:inter}
\end{equation}
These values are in rough agreement with the estimate given by Eq.~(\ref{eq:ts-crit}).
\\

Quantization may also originate from confinement along the width $W$ of the nanosheet. Although we have not calculated it here, we can estimate from
the results of Ref.~\cite{Bimin_2025} the coherence length, $\lambda_{\rm phi, 1D}$, of the 1DEG that would result assuming confinement also along the width of a 
Si nanosheet (that should now be called `nanoribbon'). Considering free-standing boundary conditions for the phonons, appropriate to a width of 12~nm, from 
Fig.~5 of that reference we see that the low-energy electron-phonon scattering rates in a 3$\times$3-cells Si nanowire are about $2 \times 10^{14}$/s 
for a `width' of 1.6~nm. Assuming that the rates scale as $1/W$ (an assumption that seems reasonable in the electrical quantum limit of a 1.6~nm thick sheet, but 
that overlooks phonon confinement along the sheet thickness), this yields a scattering rate of about $2.7 \times 10^{13}$/s in our 12~nm-wide nanosheet. This
results in a coherence length $\lambda_{\phi, {\rm 2D}} \sim$ 5.5~nm, having assumed $\upsilon_{\rm g} = \hbar \pi/(m_{\rm T} W)\sim 1.5 \times 10^7$~cm/sec, 
$m_{\rm T}\approx 0.2 \ m_0$, and $W=12$~nm. Considering that additional inelastic scattering processes, such as IPP scattering and electron-electron interactions, 
are likely to reduce even more the electron coherence length, in Si nanosheets 12~nm-wide the electrons are confined only along the `thickness' direction and not 
along the `width', thus constituting a 2DEG, {\em not} a 1DEG. This is consistent with what we can conclude by considering, instead, the 2D coherence length: From 
the scattering rates calculated here, $\approx 8 \times 10^{13}$/s for FSBCs and $\approx 3 \times 10^{13}$/s for CBCs at the SiO$_2$/HfO$_2$ interfaces, we get
$\lambda_{\phi, {\rm 2D}} \sim$ 2 or 5~nm for FSBSc or CBCs, respectively. Once again this value is smaller that the width of the nanosheet/ribbon.\\   
    
Finally, in analogy with Eq.~(\ref{eq:ts-crit}), assuming that $1/\tau_{\phi}$ is roughly independent of the sheet width for width larger than a few nm,
we can obtain an estimate of the critical width, $W_{\rm crit}$, of a 1.6~nm-thick (100) Si sheet below which the electrons constitute a 1DEG:
\begin{equation}
\frac{1}{\tau_{\phi}} \approx  \frac{\hbar \pi}{m_{\rm T} W_{\rm crit}^{2}} \ \ \ \ \ \ \mbox{or} \ \ \ \ \ \
        W_{\rm crit} = \left ( \frac{\hbar \pi}{m_{\rm T}} \tau_{\phi} \right )^{1/2} \sim 8 \ \mbox{nm} \ , 
\label{eq:W-crit}
\end{equation}
for $1/\tau_{\phi}=3 \times 10^{13}$/s. 
The fact that Eqs.~(\ref{eq:ts-crit}) and (\ref{eq:W-crit}) result in similar numbers is not unexpected, as `thickness' and `width' 
are interchangeable terms. 
\\

In summary, for (100) Si nanosheets with edges along the (100) direction, the critical phase-coherence length (width and thickness) is $\approx$~8~nm. 
In nanosheets wider or (and) thicker than this, the electrons can be considered to be a 2DEG (3DEG). For sheets with edges along the (011) 
direction, the critical width is reduced to about 6~nm. Accounting for additional inelastic scattering processes, beyond electron-phonon interactions, 
the critical length will be reduced quite significantly. For example, considering scattering with the IPPs, the scattering rates for thermal electrons shown
in Fig.~10 of Ref.~\cite{Mansoori_2026} are more than one order of magnitude larger, of the order of $10^{14}-10^{15}$/s, than what we have assumed so far. 
Therefore, also the phase-coherence length will be reduced by the same amount. Even in free-standing Si nanosheets, scattering with surface optical phonons would
have a qualitatively similar effect. Therefore, it is sensible to treat electrons in a Si nanosheet 1.6-to3~nm thick 
and 12~nm-wide as a 2DEG and nit as 1DEG.
\subsection{Electron transport} 
As a side issue of these considerations, we may also discuss whether or not we may employ a semiclassical formulation of electron transport or we need a full quantum
formulation. Indeed, once again,  it is the phase-coherence that gives us the answer. However, now we should consider the phase-coherence length projected along 
the transport direction, $\lambda^{\rm (t)}_{\phi, {\rm el}}$, since this is the quantity that  
determines whether or not the electron may `feel' the variation of the (source-to-drain) potential along the transport direction. Moreover,
this quantity varies along the channel, as electrons are heated by the applied bias. Assuming now some energy-dependent form, 
$\lambda^{\rm (t)}_{\phi, {\rm el}}(E) \sim \upsilon_{\rm d}(E) \tau^{\rm (ep)}(E)$, near the source-end of the channel $\upsilon_{\rm d}(E)\sim 10^6$~cm/s and
$1/\tau^{\rm (ep)} \approx 3 \times 10^{13}$/s - $8 \times \times 10^{13}$/s, as before. This leads to a coherence length of approximately 0.1-to-0.3~nm. 
At a high drain bias (the 'ON' state of the device), this value may be smaller for hot electrons at the drain-end of the channel: Whereas $\upsilon_{\rm d}(E)$ 
may slightly exceed 10$^7$~cm/s, $1/\tau^{\rm (ep)}(E)$ may be as high as $\sim 5 \times 10^{14}$/s at energies of a few eVs, so that
$\lambda^{\rm (t)}_{\phi, {\rm el}}(E)$ will be in the range of 0.2~nm. In either case, electrons can be assumed to loose memory of their phases along a short
portion of the channel and may be treated as semiclassical particles. This conclusion is reinforced by the fact that additional inelastic scattering
processes may reduce the phase-coherence length even more.\\  

This discussion makes us wonder if, ignoring quantum-computing applications, a quantum formulation of electronic transport at or above room temperature is ever needed.
Indeed, granting that full quantum-transport studies provide useful information about the ultimate performance limits of materials and devices, in practice
effects due to electron quantum interference have been observed experimentally only in devices based on suitable materials (such as on high-electron-mobility GaAs) 
at cryogenic temperatures as, for example, when 
detecting the Aharonov-Bohm effect~\cite{Timp_1988,Ford_1988} or quantum interference in ballistic hot-electron transistors~\cite{Hayes_1985,Heiblum_1987}. 
Obviously, quantum tunneling/interference is the operating principle of Esaki and resonant tunnel diodes~\cite{Iogansen_1964,Esaki_1973,Sollner_1983} 
tunnel FETs~\cite{Appenzeller_2004,Seabaugh_2010,Zhang_2016}, and superconductor-based SQUIDs~\cite{SQUID_1964}.  
However, we are not aware of any direct observation of the signature of quantum transport in conventional devices operating at room temperature (or above). 
To express an arguably controversial opinion, we suspect that purely ballistic transport at room temperature is a chimera: In addition to electron-phonon
scattering, irreversible Coulomb electron-electron interactions, also with electrons in metallic contacts, may prevent the onset of such a regime.
\begin{figure}[tb]
\centerline 
{\hbox{
\includegraphics[width=8.6cm]{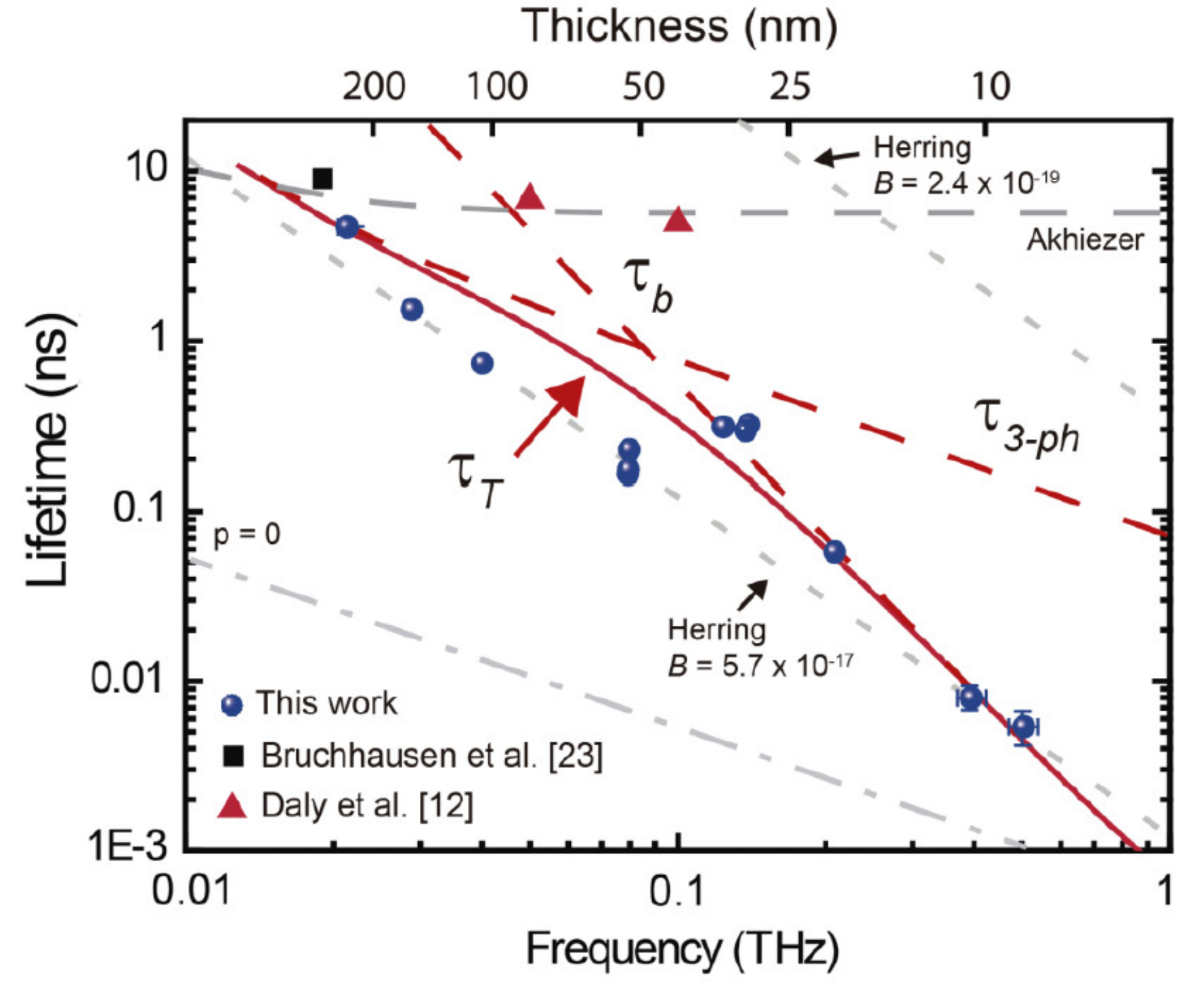}}}
\caption{Figure 3 from Ref.~\cite{Cuffe_2013} "{\it Phonon lifetime of the first-order dilatational 
         mode in free-standing silicon membranes as a function of frequency. Experimental data of free-standing silicon membranes with thickness values ranging from
         approximately 222 to 8~nm (black square [28], blue circles) and bulk silicon (red triangles [12]). The red dashed lines show the contributions to the
         finite phonon lifetime from normal three-phonon interactions $1/\tau_{3-ph}$ and boundary scattering $1/\tau_{b}$ as indicated. The total
         contribution, calculated using Matthiessen's rule $1/\tau_{T}=1/\tau_{3-ph}+1/\tau_{b}$ is shown by the solid red line labelled $1/\tau_{T}$. Other models 
         for intrinsic (grey dotted line: Herring [12,30], grey dashed line: Akhiezer [12,18]) and extrinsic (dotted-dashed grey line: Casimir limit
         $p=0$) scattering processes are shown for reference. The top axis showing thickness applies only to the experimental data presented in this Letter and 
         the extrinsic scattering processes.}" The references and reference numbers are those of the original article.
         [Figure and words in quotes/italic reprinted with permission from J.~Cuffe, O.~Ristow, E.~Ch\'{a}vez, A.~Shchepetov, P.O.~~Chapuis, F.~Alzina, M.~Hettich,
         M.~Prunnila, J.~Ahopelto, T.~Dekorsy, and C.~M.~Sotomayor Torres, Phys. Rev. Lett. {\bf 110}, 095503 (2013), \copyright2013 by the American Physical Society.]}   
\label{fig:phonon_lifetime}
\end{figure}
\section{\label{sec:Phonons}{Phonon confinement in S\lowercase{i} nanosheets}}
The discussion regarding phonon confinement is more complicated. As for electrons, we can define a phonon phase-coherence
length $\lambda_{\phi, {\rm ph}}=\upsilon_{\rm g} \tau_{\rm T}$, where $\upsilon_{\rm g}$ is the phonon group velocity and $\tau_{\rm T}$ the total inelastic lifetime 
(both quantities being, in principle functions of branch and frequency). Considering only acoustic phonons, their lifetime $\tau_{\rm T}$ is determined by 
anharmonic 3-phonon processes, $\tau_{\rm 3-ph}$, scattering with boundaries, $\tau_{\rm b}$ (most important in thin films) and, of course, by electron-phonon
interaction, $\tau^{\rm (ep)}$. Therefore, the phonon decay rate may be expressed as:
\begin{equation}
\frac{{\rm d} N_{\rm ph}}{{\rm d}t} \approx \frac{N_{\rm ph}}{\tau_{\rm 3-ph}} + \frac{N_{\rm ph}}{\tau_{\rm b}} + \frac{n_{\rm el}}{\tau^{\rm (ep)}} \ , 
\label{eq:phonon_decay}
\end{equation}   
where $1/\tau^{\rm (ep)}$ is the rate at which electrons absorb phonons.
Note that the last term is proportional to the electron density, $n_{\rm el}$, since $1/\tau^{\rm (ep)}$ already includes the phonon density via the Bose-Einstein
factor $N(\omega)$ that enters the electron-phonon matrix elements. Indeed, $n_{\rm el}$ is the bottleneck: In the absence of free electrons, the phonons cannot decay via electron-phonon processes.\\

Regarding boundary scattering, this process does not lead to a loss of phase coherence: The apparent loss of information of the phase is due 
to the average over random configurations of the rough boundary, much as it happens in the case of ionized-impurity or surface-roughness scattering for electrons. Therefore, while this process may be viewed as dissipative (that is: irreversible), it remains elastic (although accounting for multi-phonon inelastic boundary scattering has been claimed to play a role in the Kapitza thermal conductance~\cite{Hopkins_2009}); phonons do not interact with scatterers that possess internal degrees of freedom about which we lose information. Therefore, since we are interested in processes that lead to a loss of information (so, inelastic), we should
consider only $\tau_{\rm 3-ph}$ and $\tau^{\rm (ep)}$.\\

Figure~\ref{fig:phonon_lifetime} from Ref.~\cite{Cuffe_2013} shows the experimentally measured lifetime of acoustic phonons in freestanding thin Si membranes. 
The lifetime due to 3-phonon processes obviously decreases with increasing phonon frequency (as more decay channels are available to higher-frequency modes). 
However, even at the highest frequency measured, it remains large, exceeding tens or hundreds of ps. To determine the relative importance of the first and third 
terms on the right-hand side of Eq.~(\ref{eq:phonon_decay}), and so estimate $\lambda_{\phi, {\rm ph}}$, we need two pieces of information: 1. The electron-phonon
scattering rate for each single phonon branch (Fig.~6 from Ref.~\cite{Mansoori_2026} shows only the total electron-phonon scattering rates) and, 2. an estimate of the 
phonon density, 
$N_{\rm ph}$.\\
\begin{figure*}[tb]
\centerline 
{\vbox{
{\hbox{
\includegraphics[width=8.60cm]{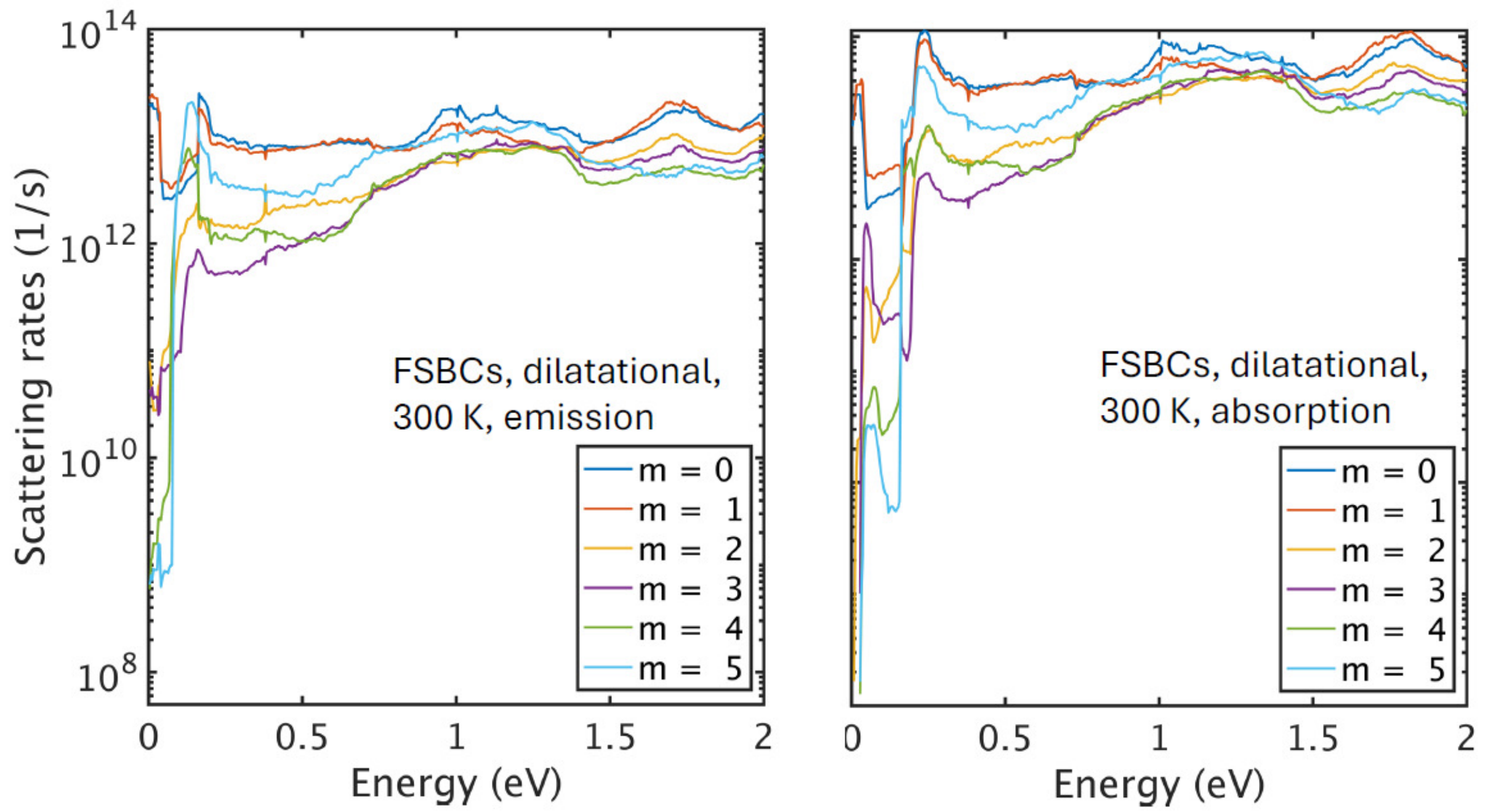}
\includegraphics[width=8.60cm]{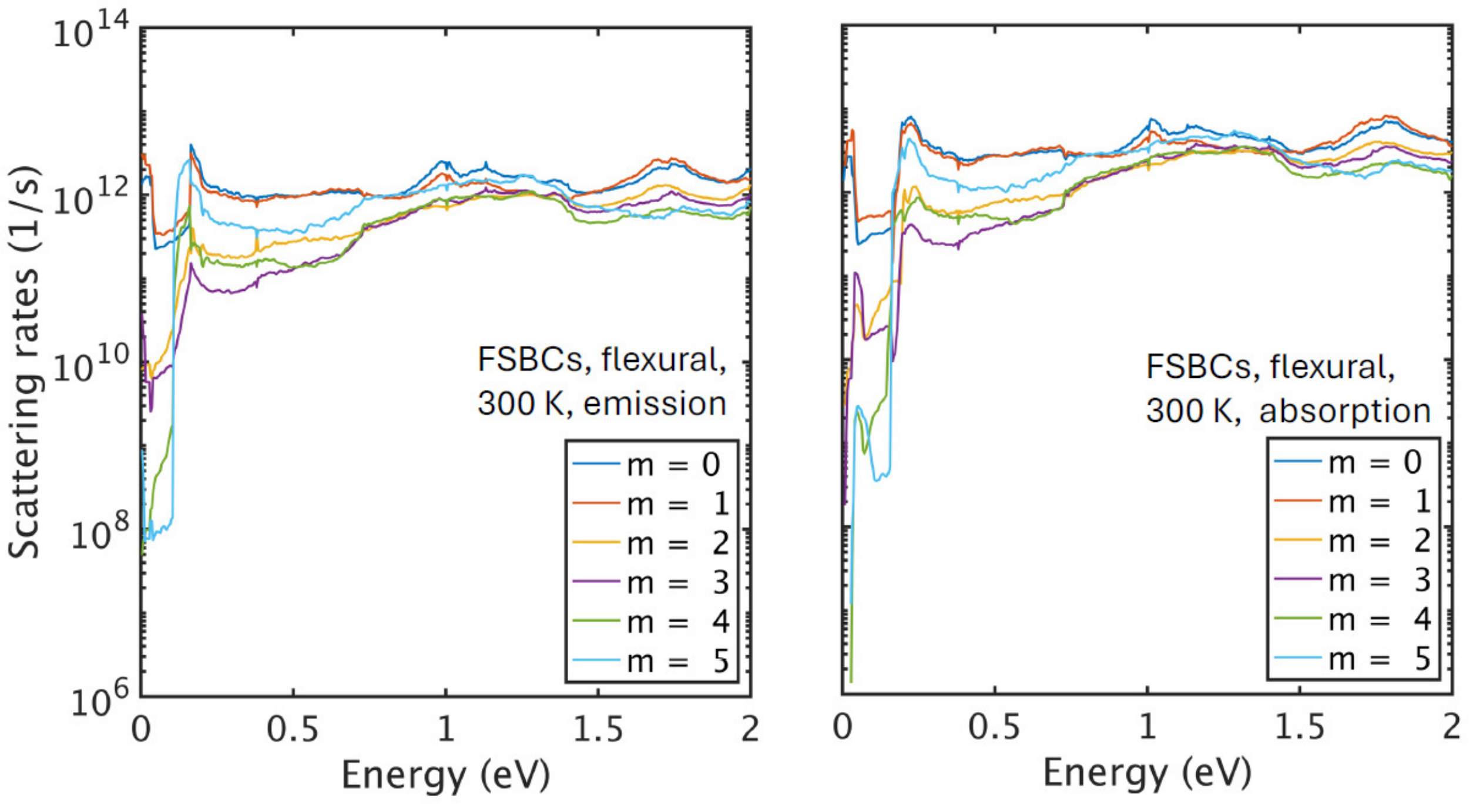}}}
{\hbox{
\includegraphics[width=8.60cm]{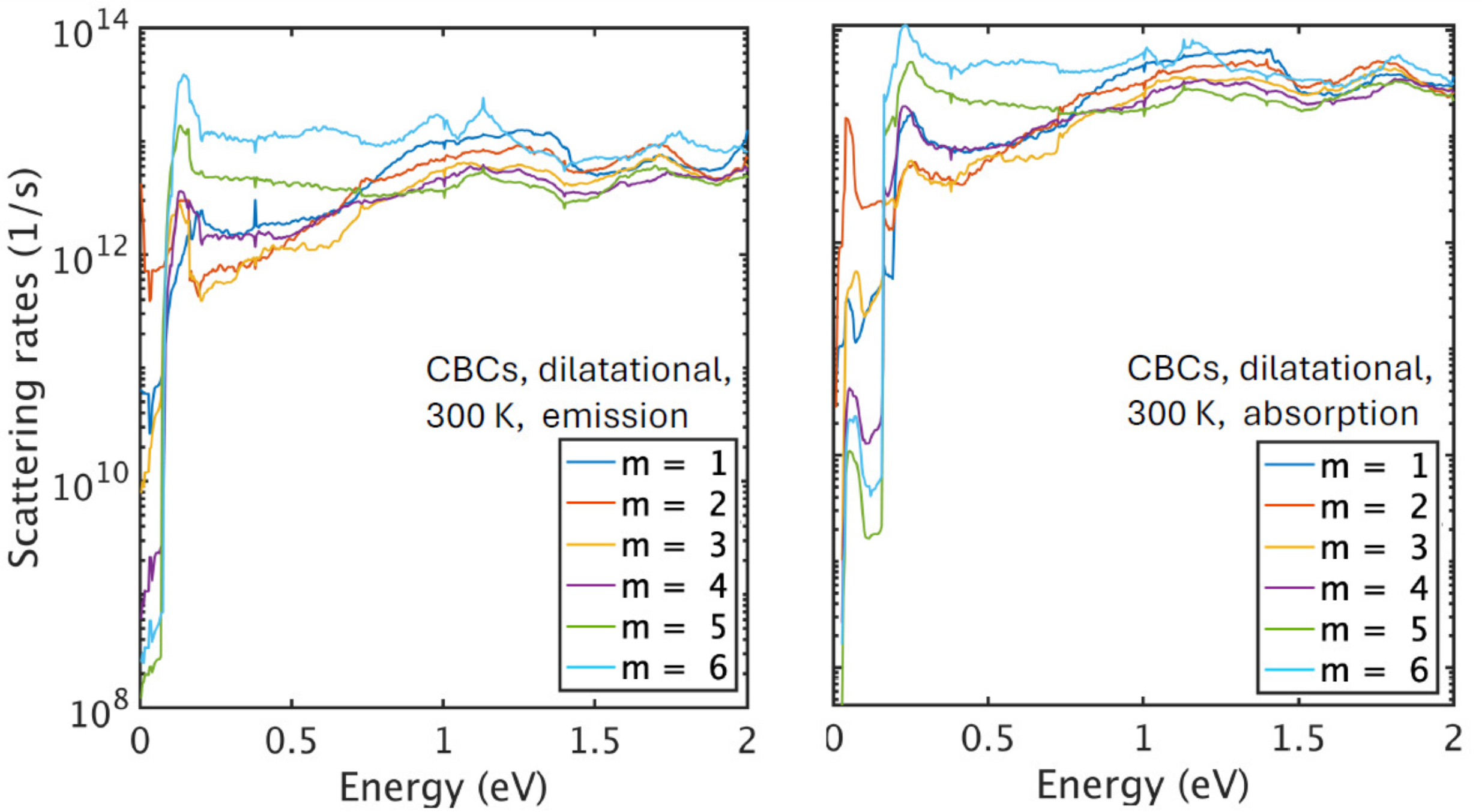}
\includegraphics[width=8.60cm]{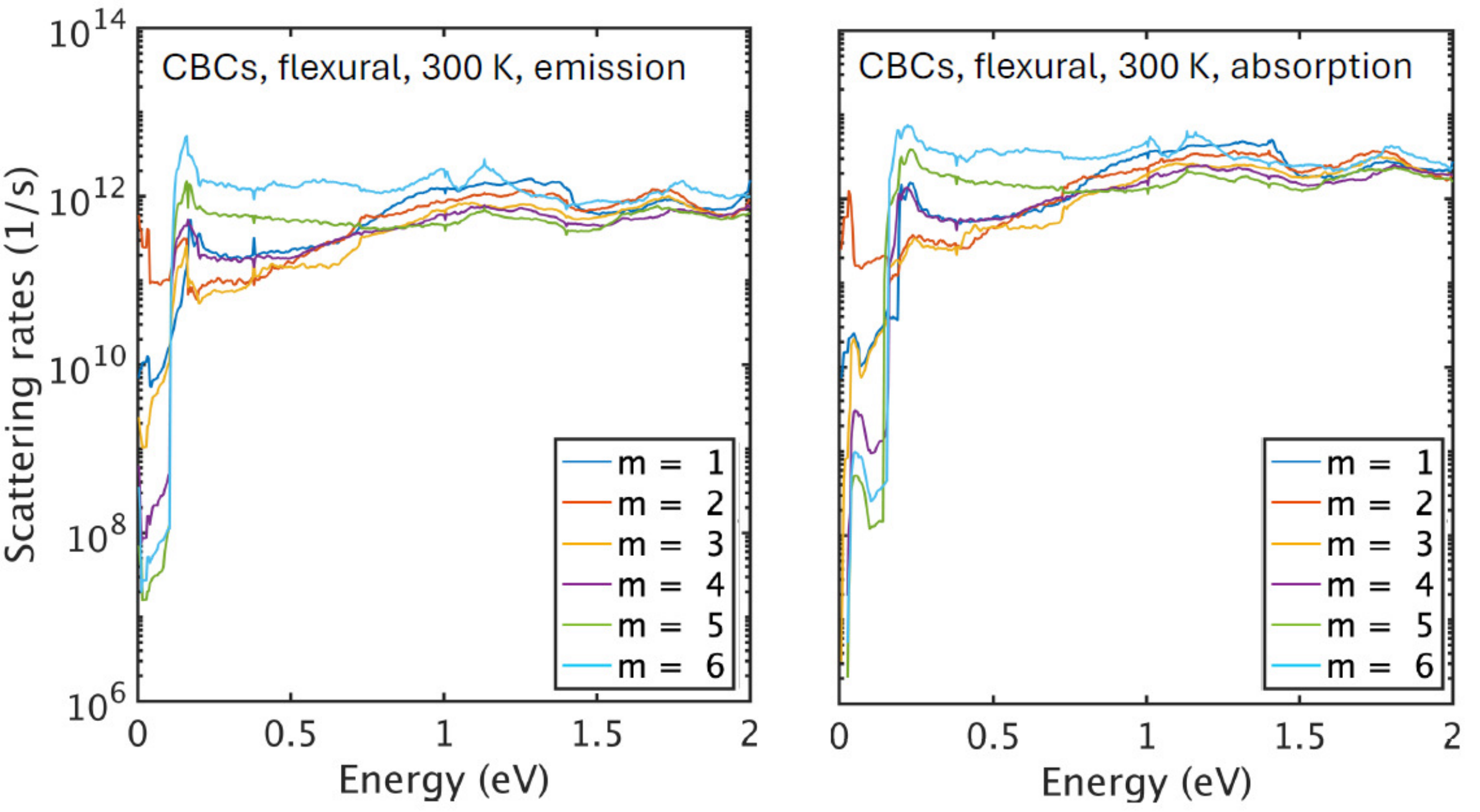}}}
}}
\caption{Calculated rates for electron emission and absorption of the confined LA (dilatational) and TA (flexural) phonons in the Si nanosheets considered
         in Ref.~\cite{Mansoori_2026}: A 1.63~nm-thin (100) Si film with top and bottom gate stacks composed by a 0.6~nm-thin SiO$_2$ and a 1.6~nm-thin
         HfO$_2$ insulating layers. The integer labels $m$ in the  legends refer to the confined vibrational modes in the nanosheet assuming clamped (CBC) or 
         free-standing (FSBC) boundary conditions at the Si/SiO$_2$ interfaces. Reference~\cite{Mansoori_2026} provides all details about the methods used to 
         obtain the data shown here.}   
\label{fig:El-ph_3-cell_LA-TA}
\end{figure*}

The first piece of information is given by Fig.~\ref{fig:El-ph_3-cell_LA-TA} that shows the electron-phonon scattering rates separately for emission and absorption of 
dilatational and flexural phonons assuming FSBCs and CBCs at the Si/SiO$_2$ interfaces, conditions that we consider as extreme cases to determine the lower and upper bounds of the phonon lifetime. 
The second piece of information, the phonon density, can be obtained following a Debye-like approach.
Considering a single acoustic phonon with dispersion $\omega(q) = c_{\rm s}q$, where $c_{\rm s}$ is the sound velocity, the bulk density of states is:
\begin{multline}
{\mathcal{D}}^{\rm (3D)}_{\rm ph}(E) = \frac{1}{(2 \pi)^3} \int {\rm d} {\bf q} \ \delta(E-\hbar c_{\rm s} q) = \\ 
                            = \frac{1}{2 \pi^2} \int {\rm d} q \ q^2 \ \delta(E-\hbar c_{\rm s} q) = \\
                            = \frac{1}{2 \pi^2} \frac{1}{(\hbar c_{\rm s})^3} \int {\rm d} E' E'^2  \delta(E-E') =
                            \frac{1}{2 \pi^2} \frac{E^2}{(\hbar c_{\rm s})^3} \ .
\label{eq:DoS_ph_3D}
\end{multline}
Therefore, the volume density of 3D acoustic phonons is:
\begin{multline}
N^{\rm (3D)}_{\rm ac} = \int_{0}^{E_{\rm BZ}} {\rm d} E \ \frac{{\mathcal{D}}^{\rm (3D)}_{\rm ph}(E)}{e^E-1} = \\
                = \frac{1}{2 \pi^2} \left ( \frac{k_{\rm B}T}{\hbar c_{\rm s}} \right )^3 \int_{0}^{T_{\rm D}/T} {\rm d} x \frac{x^2}{e^x-1} = \\
                = \frac{1}{2 \pi^2} \left ( \frac{k_{\rm B}T}{\hbar c_{\rm s}} \right )^3 D_{2} \left ( \frac{T_{\rm D}}{T} \right ) \ ,
\label{eq:Nph_ph_3D}
\end{multline}
where $E_{\rm BZ}=\hbar c_{\rm s}q_{\rm BZ}$ is the maximum phonon energy at the edge of the Brillouin zone (and we will assume $q_{\rm BZ} \sim 2\pi/a_{0}$), 
$T_{\rm D}=\hbar c_{\rm s} q_{\rm BZ}/k_{\rm B}$ is approximately the Debye temperature ({\it i.e.}, larger than the Debye temperature by a factor of 
$(4\pi/3)^{1/3} \approx 1.61$ originating from having approximated the Brillouin zone as a sphere of radius $2\pi/a_0$ rather than
$(6\pi^{2})^{1/3}/a_0$, as required to have the volume of a cube with sides $2\pi/a_0$), and $D_2(y)=\int_{0}^{y} {\rm d} x \ x^2/(e^x-1)$ is the second 
Debye integral. Converting this volume density to a surface density in the nanosheet, the surface density is:
\begin{equation}
N^{\rm (2D)}_{\rm ac} =  t_{\rm s} N^{\rm (3D)}_{\rm ac} =
                 \frac{t_{\rm s}}{2 \pi^2} \left ( \frac{k_{\rm B}T}{\hbar c_{\rm s}} \right )^3 D_{2} \left ( \frac{T_{\rm D}}{T} \right ) \ ,
\label{eq:Nph_ph_2D_from_3D}
\end{equation}
In our case, assuming $c_{\rm LA} \approx 10 \times 10^5$~cm/s, $c_{\rm TA} \approx 5 \times 10^5$~cm/s, and $t_{\rm s} = 3 a_{0}$,  we obtain
$N^{\rm (2D)}_{\rm LA} \approx 7.5 \times 10^{14}$~cm$^{-2}$ and $N^{\rm (2D)}_{\rm TA} \approx 2.7 \times 10^{15}$~cm$^{-2}$ at 300~K. 
Using, instead, the `correct' Debye value $q_{\rm BZ} = (6 \pi^2)^{1/3}/a_0$, these values become 
$N^{\rm (2D)}_{\rm LA} \approx 4.4 \times 10^{14}$~cm$^{-2}$ and $N^{\rm (2D)}_{\rm TA} \approx 2.1 \times 10^{15}$~cm$^{-2}$
These values should be viewed as lower bounds, since the phonon dispersion `flattens' with increasing $q$, so that the factor $e^x$ in the denominator of the integrand in $D_2(y)$ is
overestimated by this Debye-like model.\\ 

Assuming, instead, acoustic phonons confined along the $z$ direction, the 2D phonon density of states for a single branch is:
\begin{multline}
{\mathcal{D}}^{\rm (2D)}_{\rm ph}(E) = \frac{1}{(2 \pi)^2} \int {\rm d} {\bf Q} \ \delta(E-\hbar c_{\rm s} Q) =  \\
                            = \frac{1}{2 \pi} \int {\rm d} Q \ Q \ \delta(E-\hbar c_{\rm s} Q) = \\
                            = \frac{1}{2 \pi} \frac{1}{(\hbar c_{\rm s})^2} \int {\rm d} E' E'  \delta(E-E') =
                            \frac{1}{2 \pi} \frac{E}{(\hbar c_{\rm s})^2} \ ,
\label{eq:DoS_ph_2D}
\end{multline}
which implies for the phonon density:
\begin{multline}
N^{\rm (2D)}_{\rm ac} = \int_{0}^{E_{\rm BZ}} {\rm d} E \ \frac{{\mathcal{D}}^{\rm (2D)}_{\rm ph}(E)}{e^E-1} = \\
                = \frac{1}{2 \pi} \left ( \frac{k_{\rm B}T}{\hbar c_{\rm s}} \right )^2 \int_{0}^{T_{\rm D}/T} {\rm d} x \frac{x}{e^x-1} = \\
                = \frac{1}{2 \pi} \left ( \frac{k_{\rm B}T}{\hbar c_{\rm s}} \right )^2 D_{1} \left ( \frac{T_{\rm D}}{T} \right ) \ ,                 
\label{eq:Nph_ph_2D}
\end{multline}
where $D_1(y)=\int_{0}^{y} {\rm d} x \ x/(e^x-1)$ is the first Debye integral. In this case, we obtain
$N^{\rm (2D)}_{\rm LA} \approx 3.6 \times 10^{14}$~cm$^{-2}$ and $N^{\rm (2D)}_{\rm TA} \approx 1.2 \times 10^{15}$~cm$^{-2}$. 
Assuming, instead, $q_{\rm BZ}= (2/\pi)^{1/2}/a_0$ (as required to obtain the correct area of the 2D Brillouin zone), these values become
$N^{\rm (2D)}_{\rm LA} \approx 3.3 \times 10^{14}$~cm$^{-2}$ and $N^{\rm (2D)}_{\rm TA} \approx 1.06 \times 10^{15}$~cm$^{-2}$.
Accounting for the 6 quantized branches (with a higher DoS but at higher energy), these values are comparable to those obtained assuming bulk (3D) phonons.\\

Having obtained this information, we can assume $N_{\rm dil} \approx 5 \times 10^{14}$~ cm$^{-2}$, $N_{\rm flex} \approx 1.5 \times 10^{15}$~ cm$^{-2}$,
$1/\tau^{(ep)} \approx 10^{12}$/s (CBCs) and $\approx 5 \times 10^{13}$/s (FSBCs) and 
$\tau_{\rm 3-ph} \sim$ 10~ps-to-1~ns. Therefore, in Eq.~(\ref{eq:phonon_decay}), ${\rm d}N_{\rm ph}/\tau_{\rm 3-ph}$ can range from 
$5 \times 10^{23}$~cm$^{-3}$s$^{-1}$ to $1.5 \times 10^{26}$~cm$^{-3}$s$^{-1}$, whereas $n_{\rm el}/\tau^{\rm (ep)}$ can range from $10^{24}$~cm$^{-3}$s$^{-1}$ 
to $10^{25}$~cm$^{-3}$s$^{-1}$ (for $n_{\rm el}$ in the range $10^{12}-10^{13}$~cm$^{-2}$). Defining an effective phase-coherence time $\tau_{\rm eff}$ 
as $\tau^{-1}_{\rm eff} = \tau^{-1}_{\rm 3-ph}+(n_{\rm el}/N_{\rm ph})\tau^{\rm (ep)-1}$, we can estimate an effective phase-coherence length for long-wavelength acoustic phonons $\lambda_{\phi, {\rm ph}} = c_{\rm s} \tau_{\rm eff}$. This can range from 50~nm to as much as 10~$\mu$m. Therefore, 
decay via anharmonic processes is likely to be the process that controls the phonon lifetime~\cite{bnote2}. At the higher temperature at which devices operate, 
$80^{\circ} \sim 350$~K, this process is likely to dominate even at an electron sheet density of $10^{13}$~cm$^{-2}$, since the phonon density increases 
as T$^{2}$ (ignoring a weaker $T$-dependence of $D_{1}(T_{\rm D}/T)$ for $T \lesssim T_{\rm D} \sim$ 400-700~K) 
and $1/\tau_{3-ph} \sim T^{\gamma}$ where $\gamma>2$~\cite{Guthrie_1966}, 
whereas $1/\tau_{\rm}$ increases as $T$ (for electron scattering with long-wavelength acoustic-phonons). Shorter-wavelength acoustic phonons and 
optical phonons certainly will have a shorter lifetime and coherence length. Still, even at 300~K, all phonons will be confined along the $z$ direction for
films thinner than about 10~nm.\\

More interestingly, though, long-wavelength acoustic phonons may 
be confined also along the $y$ direction for nanosheets with a width smaller than 50~nm, rendering them structures in which some phonon branches will be 
one-dimensional, as in the case of nanowires and nanoribbons. Treating acoustic phonons in such a situation (long-wavelength 1D phonons, shorter-wavelength
2D phonons) may resemble the treatment of electrons in a quantum well with finite potential barriers of height $\Phi_{\rm B}$: Low-energy electrons form a 2DEG 
on the $(x,y)$ plane, but as the energy of the eigenstates $\ket{\mu}$, 
$E_{\mu}$, approaches $\Phi_{\rm B}$, their dispersion begins to acquire a dependence on $k_{z}$ and, 
for $E_{\mu} > \Phi_{\rm B}$, they become fully delocalized bulk particles with a dispersion that approaches the free-electron dispersion at higher energies.
Thus, we may treat acoustic phonons as confined by an effective potential with `barriers' (whose height is dictated by the elastic mismatch at the Si/insulator
interfaces) at $y=0$ and $y=W_{\rm s}$ (where $y=W_{\rm s}$ is the width of the nanosheet).\\

For long-wavelength phonons these conclusions do not change, even if we were willing to consider boundary scattering as inelastic, since this process is largely
ineffective at long wavelength~\cite{Cuffe_2013}. Shorter-wavelength phonons, as well as optical phonons, are likely to have a much shorter lifetime anyway, 
of the order of several ps, therefore it is unlikely that they will be confined `laterally' even for nanosheets as narrow as 10~nm. Moreover, as mentioned
above, CBCs at the SiO$_2$/Si interfaces may be considered appropriate, since the optical `mismatch' between these two materials is much larger than their acoustic
mismatch, as given by the Young modulus.

\section{\label{sec:conclusion}{Conclusions}}
In order to assess the electronic and vibrational dimensionality of a semiconductor nanostructure, we have argued that the electron and phonon coherence length
(their inelastic mean free path) sets the length scale below which quantum confinement comes into play. We have considered specifically the example of Si 
nanosheets at room temperature and have estimated that the critical length below which electrons are subject to quantum confinement is of the order of (or smaller
than) 8~nm at 300~K, considering energy losses to phonons and remote phonons in gated structures. The situation is much more complicated when dealing with phonons. Considering their coherence length as determined by scattering with electrons and anharmonic three-phonon processes, we have argued that short wavelength acoustic 
and optical phonons may be confined only by structures as small as 10~nm. On the contrary, long-wavelength acoustic phonons may be affected by geometrical 
confinement even at the $\mu$m length scale. Whereas these estimates apply to Si nanosheets at 300~K, the qualitative physical picture should be valid for a variety
of semiconductors.    

\acknowledgments
This work has been supported by the Taiwan Semiconductor Manufacturing Company Ltd. (TSMC). We also thank the Texas Advanced Computing Center (TACC) for having provided the computing resources required to perform this study and Texas Instruments for the Endowment that has provided additional financial support.

\bibliography{paper_ref_titles}

\end{document}